\documentclass[aps,prb,preprint,groupedaddress,showpacs,showkeys,amsmath]{revtex4-1}
\usepackage{graphicx}
\usepackage{dcolumn}
\usepackage{bm}
\usepackage{color}
\begin{document}

\title{Systematic study of finite-size effects in quantum Monte Carlo calculations of real metallic systems}

\author{Sam Azadi}\email{s.azadi@imperial.ac.uk}
\affiliation{Department of Physics, Imperial College London, Exhibition Road, 
London SW7 2AZ, United Kingdom}
\author{W. M. C. Foulkes}
\affiliation{Department of Physics, Imperial College London, Exhibition Road, 
London SW7 2AZ, United Kingdom}
\date{\today}

\begin{abstract}
  We present a systematic and comprehensive study of finite-size effects
  in diffusion quantum Monte Carlo calculations of metals. Several
  previously introduced schemes for correcting finite-size errors are
  compared for accuracy and efficiency and practical improvements are
  introduced. In particular, we test a simple but efficient method of
  finite-size correction based on an accurate combination of twist
  averaging and density functional theory. Our diffusion quantum Monte
  Carlo results for lithium and aluminum, as examples of metallic
  systems, demonstrate excellent agreement between all of the approaches
  considered.
\end{abstract}

\maketitle

\section {Introduction}

Density functional theory (DFT) has dominated atomic-scale materials
modeling for the past three decades and will continue to be extremely
important. However, there are plenty of problems where the accuracy of
DFT falls short of requirements. There are clear reasons to expect DFT
to struggle in strongly-correlated solids, but these are not the only
systems for which DFT is insufficient. Take, for example, the problem of
distinguishing between molecular crystal phases and competing low-energy
polymorphs. Even in relatively simple molecular solids such as
crystalline benzene and its polymorphs under pressure, the energy
differences of interest are less than a few kJ/mol.  The most successful
calculations based on DFT are only reliable to $\sim$10
kJ/mol,\cite{benzene1} and it has recently been shown that the use of
{\it ab initio} many-electron wavefunction methods, such as quantum
Monte Carlo (QMC), is essential to tackle this problem
successfully.\cite{benzene2} Unlike DFT, many-electron wavefunction
methods can, in principle, be improved systematically until the required
convergence is obtained.

Another important example is the adsorption of molecules on surfaces,
where DFT is sometimes unable to give predictions of useful
accuracy. DFT values of surface formation energies of simple
paradigmatic materials such as silicon and magnesium oxide depend
strongly on the assumed exchange-correlation functional, and there is
usually no way of knowing in advance which functional to trust.  The
well known problem of calculating electronic band gaps could also be
mentioned.

The urgent practical need to go beyond DFT in these and other areas is
driving current efforts to develop more accurate methods.  There is
abundant evidence that there are large classes of problems for which QMC
techniques, in particular diffusion Monte Carlo (DMC), are considerably
more accurate that DFT.\cite{Hamdani,cox,santra,marchi} Just recently, using
QMC methods, chemically accurate ionization potentials have been
obtained for the first-row transition-metal atoms with a mean absolute
error of only 0.126 kcal/mol.\cite{Thomas} By combining high accuracy
with good efficiency and scalability, QMC methods promise to bring high
accuracy to computational materials science as a matter of
routine.\cite{Booth}

QMC simulations of extended systems are carried out using finite
simulation cells subject to periodic boundary conditions. Practical and
computational constraints restrict the size of the simulation cell and
so introduce finite-size (FS) errors, which can be large and are one of
the main problems holding back the application of accurate QMC
techniques to solids.\cite{Matthew1,Neil} Quantifying and correcting
these errors is an essential part of all QMC simulations of extended
systems, particularly when high accuracy is required.

FS errors affect independent-particle approaches such as DFT as well as
many-body approaches such as QMC. For calculations of perfect crystals,
DFT FS errors can be reduced simply by improving the accuracy of the
Brillouin zone integration, although the errors arise when periodic
supercells are used to model aperiodic systems are less easily dealt
with. Independent-particle FS errors also affect many-body calculations
of perfect crystals\cite{Rajagopal} and can be reduced using twist
averaging, which is the many-electron equivalent of Brillouin zone
integration. Even after twist averaging, however, ``many-body'' FS
errors with no independent-particle analogues remain.

Another important contribution to the FS error in many-body simulations
of extended systems arises from the treatment of the potential
energy. The $1/r$ Coulomb interaction is inconsistent with the
periodicity of the simulation cell and has to be replaced by the Ewald
interaction, which is the Green's function of Poisson's equation subject
to periodic boundary conditions.  Unlike the Coulomb interaction, the
Ewald interaction depends on the size and shape of the simulation cell,
leading to additional finite-size errors.\cite{Fraser} One approach to
circumventing this problem is to use a different periodic function, the
``model periodic Coulomb'' (MPC) interaction, in place of the Ewald
interaction.\cite{Fraser,Williamson} Variational quantum Monte Carlo
(VMC) simulations using the MPC interaction suffer from smaller FS
effects than simulations using the standard Ewald interaction. It has
also been shown that using the MPC interaction reduces the FS errors in
DMC calculations of ground and excited states.\cite{Kent}

A drawback of the MPC approach is that it only reduces FS errors arising
from the use of the Ewald interaction.  The charge density and
exchange-correlation hole in a finite simulation cell are often very
similar to those of an infinite solid, so the errors in the Coulomb
energy are indeed primarily due to the errors in the interaction; but
the imposition of periodic boundary conditions also affects the
many-electron wavefunction and thus the electronic kinetic energy. The
non-interacting part of the FS error in the kinetic energy may be
eliminated by twist averaging, but the many-body contributions are not
negligible.

Under the assumption that the low-$k$ behaviour of the structure factor
is independent of the choice of simulation cell, Chiesa \emph{et
  al.}\cite{Chiesa} proposed a method to estimate the many-body
contributions to the FS errors in both the potential and kinetic
energies without abandoning the Ewald interaction.  Employing this
correction, which is based on the random-phase approximation at long
wavelength, one can calculate FS corrections within a single simulation.

Another approach to the treatment of many-body finite-size errors is
provided by the Kwee-Zhang-Krakauer (KZK) functional,\cite{kzk} which
adds a correction computed from the difference between the DFT energy
evaluated using the local density approximation (LDA) for an infinite
system and the DFT energy evaluated using a modified LDA specifically
designed to reproduce the total energy of the finite simulation cell,
including FS errors. Both the standard LDA and the KZK LDA are
parameterized on the basis of DMC simulations of cells of uniform
electron gas subject to periodic boundary conditions, but the standard
LDA uses DMC energies that have been extrapolated to infinite cell size
while the KZK LDA does not.

In this paper, we systematically study the problem of eliminating FS
errors from QMC calculations of real metallic systems, taking lithium
and aluminum as examples.  We analyze twist-averaged DMC energies
obtained using the Ewald interaction and the MPC
interaction,\cite{Fraser,Williamson} and finite-size corrections based
on the Chiesa formalism\cite{Chiesa} and the KZK functional.\cite{kzk}
We also investigate DFT-based corrections designed to improve
imperfectly twist-averaged results and consider how best to combine the
use of twist-averaged boundary conditions\cite{twistav} with the KZK
functional.\cite{kzk}

\section {Computational Details}
\subsection{Diffusion quantum Monte Carlo calculations}

The diffusion quantum Monte Carlo method is a stochastic technique for
obtaining the ground-state energy of a many-electron system. DMC has
been described in many previous papers\cite{Matthew1, Neil} and will not
be discussed in detail here, but since this work is focused on technical
aspects of DMC simulations we start with a brief explanation.

The DMC algorithm solves the imaginary-time Schr\"{o}dinger equation,
\begin{equation}
  \frac{\partial \Psi(\mathbf{R}, \tau )}{\partial\tau} = \frac{1}{2}
  \sum^{N_e}_{i=1} \nabla_{\mathbf{r}_i}^2 \Psi(\mathbf{R}, \tau )
  - (V(\mathbf{R})-E_{T})\Psi(\mathbf{R},\tau),
\end{equation}
where $\mathbf{R} = (\mathbf{r}_1,\mathbf{r}_2,\ldots,\mathbf{r}_{N_e})$
is a $3N_e$-dimensional vector defining the positions of all $N_e$
electrons in the simulation cell, $\tau$ is the imaginary time (a real
variable despite its name), $V(\mathbf{R})$ is the potential energy
including electron-electron interactions, and $E_T$ is a constant energy
offset. (We work in Hartree atomic units, where the numerical values of
$\hbar$, $e$, $m_e$, and $4\pi\varepsilon_0$ are all equal to 1.) The
imaginary-time Schr\"{o}dinger equation resembles a $3N_e$-dimensional
diffusion equation with diffusion constant $D = 1/2$.  The potential
energy term causes the diffusers to ``branch'' (multiply or die out) at
a position-dependent rate. The wavefunction $\Psi(\mathbf{R},\tau)$ is
the number density of diffusers, which are normally known as walkers or
configurations and are points in the $3N_e$-dimensional configuration
space, not individual electrons. The DMC algorithm uses this simple
physical interpretation to simulate the imaginary-time evolution of the
wavefunction using a finite population of diffusing and branching
walkers.

Solving the imaginary-time Schr\"{o}dinger equation is useful because it
projects out the ground state as $\tau \rightarrow \infty$.  If the
initial wavefunction is expanded as a linear combination of energy
eigenfunctions, $\Psi(\tau=0) = \sum_{i} c_i \Psi_i$, the solution of
the imaginary-time Schr\"{o}dinger equation $\partial\Psi/\partial\tau =
-(\skew3\hat{H}-E_T)\Psi$ is
\begin{equation}
  \Psi(\tau) = \sum_i c_i e^{-(E_i - E_T)\tau} \Psi_i.
\end{equation}
Thus, as long as $c_0 \neq 0$, the wavefunction $\Psi(\tau)$ becomes
proportional to $\Psi_0$ as $\tau \rightarrow \infty$. By gradually
adjusting $E_T$ to maintain the normalization of the solution in the
large $\tau$ limit, we can find the ground-state energy $E_0$.

An obvious difficulty with this approach is that the wavefunction
$\Psi(\mathbf{R},\tau)$, which is not necessarily positive, is
interpreted as a walker density, which must be positive. In fact, a
naive application of the DMC algorithm to a many-electron system yields
a totally symmetric many-boson ground state of no physical interest. The
fixed-node approximation introduces a trial many-electron ground-state
wavefunction, $\Psi_T(\mathbf{R})$, and forbids walker moves that cause
$\Psi_T$ to change sign. As long as $\Psi_T$ is properly antisymmetric,
this is sufficient to ensure that a fermionic solution is obtained. It
may be shown \cite{Reynolds,Matthew1} that energies calculated within
the fixed-node approximation are variational: the result is greater than
or equal to the many-fermion ground-state energy and tends to the exact
energy as the $(3N_e-1)$-dimensional nodal surface on which $\Psi_T=0$
tends to the ground-state nodal surface. The fixed-node approximation is
required for DMC simulations of large systems but is the only
fundamental limitation of the method. Other approximations, such as the
use of a finite time-step or the representation of ions by
pseudopotentials, can be made negligible or avoided given sufficient
computer time.  Fixed-node DMC energies are in most cases comparable in
accuracy to energies calculated using the CCSD(T) method,\cite{Helgaker}
which is often known as the ``gold standard'' of quantum chemistry.

The diffusion/branching simulation described above is unstable in
practice because the potential energy $V(\mathbf{R})$ diverges whenever
electrons approach nuclei or each other, leading to uncontrollable
branching. This problem can be overcome using an importance-sampling
technique. The imaginary-time Schr\"{o}dinger equation is re-expressed
in terms of the quantity $f(\mathbf{R},\tau) =
\Psi_T(\mathbf{R})\Psi(\mathbf{R},\tau)$ to obtain
\begin{equation}
  \frac{\partial f(\mathbf{R},\tau)}{\partial t} =
  \frac{1}{2}\nabla_{\mathbf{R}}^2 f(\mathbf{R},\tau) -
  \bm{\nabla}_{\mathbf{R}}\cdot \left [ \mathbf{v}(\mathbf{R})
    f(\mathbf{R},\tau) \right ] -
  [E_L(\mathbf{R}) - E_T]f(\mathbf{R},\tau),
\end{equation}
where $\bm{\nabla}_{\mathbf{R}} = (\bm{\nabla}_{\mathbf{r}_1},
\bm{\nabla}_{\mathbf{r}_2},\ldots, \bm{\nabla}_{\mathbf{r}_{N_e}})$ is the
$3N_e$-dimensional gradient operator, $\nabla_{\mathbf{R}}^2 =
\bm{\nabla}_{\mathbf{R}}\cdot\bm{\nabla}_{\mathbf{R}}$ is the
corresponding Laplacian, $\mathbf{v}(\mathbf{R}) =
\bm{\nabla}_{\mathbf{R}}\ln|\Psi_T(\mathbf{R})|$ is the $3N_e$-dimensional
drift velocity vector, and $E_L(\mathbf{R}) = (1/\Psi_T(\mathbf{R}))
\hat{H}\Psi_T(\mathbf{R})$ is the local energy.  The importance-sampled
imaginary-time Schr\"{o}dinger equation describes a diffusion process
similar to that discussed above, except that the walkers now drift with
velocity $\mathbf{v}(\mathbf{R})$ as well as diffusing and
branching. The branching rate is determined by the shifted local energy
$E_L(\mathbf{R})-E_T$ instead of the shifted potential energy
$V(\mathbf{R})-E_T$. If the trial function is a good approximation to
the ground state, the local energy is a smooth function of $\mathbf{R}$
and the numerical difficulties caused by divergences in $V(\mathbf{R})$
are avoided.  The fixed-node approximation is imposed by rejecting
walker moves that change the sign of $\Psi_T(\mathbf{R})$.

Our DMC simulations used the CASINO QMC code \cite{casino} and a trial
function of Slater-Jastrow (SJ) form,
\begin{equation}
  \Psi_{T}(\mathbf{R})=\exp[J(\mathbf{R})]
  \det[\psi_{n}(\mathbf{r}_i^{\uparrow})]
  \det[\psi_{n}(\mathbf{r}_j^{\downarrow})],
\label{eq6}
\end{equation}
where $\mathbf{r}_i^{\uparrow}$ is the position of the i'th spin-up
electron, $\mathbf{r}_j^{\downarrow}$ is the position of the j'th
spin-down electron, $\exp[J(\mathbf{R})]$ is the Jastrow factor, and
$\det[\psi_{n}(\mathbf{r}_i^{\uparrow})]$ and
$\det[\psi_{n}(\mathbf{r}_j^{\downarrow})]$ are Slater determinants of
spin-up and spin-down one-electron orbitals.  These orbitals were
obtained from DFT calculations using the plane-wave-based Quantum
Espresso code.\cite{QS} A norm-conserving pseudopotential constructed
within DFT using the Perdew-Zunger parameterization of the local density
approximation\cite{PZ} was employed. We chose a very large basis-set
cut-off of 300 Ry to guarantee converge to the complete basis-set
limit.\cite{sam} The one-electron orbitals, originally expressed as
linear combinations of plane waves, were transformed into a blip
polynomial basis for efficiency.\cite{blip} The Jastrow function
$J(\mathbf{R})$ consisted of polynomial one-body electron-nucleus (en)
and two-body electron-electron (ee) terms, the parameters of which were
optimized by variance minimization at the variational Monte Carlo (VMC)
level.\cite{varmin1,varmin2}

\subsection{Finite-size errors and correction methods}
\label{subsec:fse}

QMC FS errors are conventionally separated into one-body and many-body
contributions.  One-body (independent-particle) errors arise from the
non-interacting kinetic, potential and Hartree energies and include
shell-filling effects. These errors can be removed by twist averaging.
\cite{twistav} Many-body errors arise from the effects of exchange and
correlation on the Coulomb and kinetic energies and are not removed by
twist averaging.  As explained in the introduction, various techniques
may be used to reduce or cancel these errors, but none is entirely
successful and care is required. The oldest approach is extrapolation,
which remains useful. The use of the modified periodic Coulomb
interaction \cite{Fraser,Williamson,Kent} reduces the Coulomb errors but
not the kinetic energy errors and must therefore be combined with other
techniques.  The LDA-based Kwee-Zhang-Krakauer (KZK) approach \cite{kzk}
applies corrections obtained from DFT calculations carried out using a
modified exchange-correlation functional explicitly designed to mimic
the DMC many-body errors.

To remove single-particle errors and eliminate shell effects in the
kinetic energies of metallic systems, we use twist-averaged boundary
conditions.\cite{twistav} A twist $\mathbf{k}_s$ is imposed by insisting
that the many-electron wavefunction $\Psi_{\mathbf{k}_s}$ obeys the
Bloch boundary condition
\begin{equation}
  \Psi_{\mathbf{k}_s}(\mathbf{r}_1, ..., \mathbf{r}_i + \mathbf{L},...,
  \mathbf{r}_{N_e}) = \exp(i\mathbf{k}_s\cdot\mathbf{L}) 
  \Psi_{\mathbf{k}_s}(\mathbf{r}_1, ...,
  \mathbf{r}_i,..., \mathbf{r}_{N_e})
\end{equation}
for all electrons $i$, where $\mathbf{L}$ is any simulation-cell lattice
vector. Expectation values of observables are obtained by averaging over
twist vectors $\mathbf{k}_s$ uniformly distributed over the
simulation-cell Brillouin zone:
\begin{equation}
  \langle \widehat{O}\rangle = \frac{1}{N_{\rm twist}}
  \sum_{\mathbf{k}_s} \langle \Psi_{\mathbf{k}_s} \vert \widehat{O}\vert
  \Psi_{\mathbf{k}_s}\rangle.
\end{equation}
The twists can be chosen from a uniform Monkhorst-Pack
grid,\cite{Monkhorst} preferably offset from $\Gamma$, or can be chosen
randomly, as in this work.  The number of twists should be as large as
computational resources allow.

As the twist $\mathbf{k}_s$ varies, the energies of some of the
one-electron states appearing in the Slater determinants may cross the
Fermi level. In the canonical approach to twist averaging, the electron
number is kept constant and the Fermi level is allowed to vary with
twist. This makes the twist-averaged total energy slightly too
large,\cite{twistav, Neil} but the bias reduces as the simulation-cell
increases in size and is normally negligible. 
In the grand canonical
approach to twist averaging, the Fermi energy is fixed and the electron
number is allowed to vary with $\mathbf{k}_s$. As was demonstrated in
Ref.~\onlinecite{Neil}, energies obtained using grand-canonical twist
averaging exhibit much larger fluctuations than energies obtained using
canonical twist averaging with the same number of twists. Furthermore,
these fluctuations die away very slowly as the number of twists is
increased. To save effort, applications of QMC to real systems normally
use the smallest number of twists possible, so canonical twist averaging
is preferable despite the fact that it suffers from small systematic
errors. This work only considers twist averaging within the canonical
ensemble.

In metallic systems, even when substantial computational resources are
expended, the FS errors due to incomplete twist averaging are
substantial. We therefore define an incomplete-twist-averaging
correction as follows:
\begin{equation}
  \Delta_{BZ} = E^{DFT} (\infty) - E^{DFT}_{TAV} (L),
  \label{deltaBZ}
\end{equation}
where $E^{DFT} (\infty)$ is the DFT energy computed using a fully
converged $k$-point mesh and
\begin{equation}
  E^{DFT}_{TAV}(L) = \frac {1}{N_{\rm twist}} \sum_{\mathbf{k}_s}
  E^{DFT}(L,\mathbf{k}_s) 
\end{equation}
is the twist-averaged DFT energy obtained using the same simulation cell
and set of twists as the DMC simulation.  The incomplete-twist-averaging
correction tends to zero as the DMC twist averaging is improved and
works well if the independent-particle finite-size errors are well
approximated by their DFT equivalents.  In practice, this approach
allows accurate results to be obtained with surprisingly small sets of
DMC twists, even in metals.

We analyze three different methods for correcting the many-body FS
errors in DMC results.  Two of these use the structure-factor-based
corrections proposed by Chiesa \emph{et al.}\ \cite{Chiesa,Neil} The
first employs the standard Ewald form of the periodic Coulomb
interaction and Chiesa corrections for both the kinetic and potential
energies; the second uses the MPC\cite{Fraser,Williamson,Kent} to deal
with the Coulomb errors and a Chiesa correction for the kinetic energy
only. Results obtained with both of these methods are expected to be
similar in quality.\cite{Neil}

The third FS-correction method considered here is the KZK
approach,\cite{kzk} which uses a system-size-dependent local density
approximation fitted to the results of DMC simulations of \emph{finite}
cubic simulation cells of uniform electron gas. DFT energies calculated
using the KZK functional incorporate DMC FS errors within an
approximation analogous to the LDA. To estimate the FS error in the DMC
total energy of a given simulation cell, the DFT total energy of exactly
the same simulation cell is calculated using the KZK functional. The
difference between this value and the DFT energy of an infinite
simulation cell calculated using the standard LDA provides an estimate
of the DMC FS error. The KZK functional was not originally combined with
twist averaging, but the combination is easy to implement (see below)
and very successful.

In the following we explain how to combine twist averaging and KZK
corrections.  In general, any FS correction can be written as
\begin{equation}
  \Delta E^{FS}(L) = \Delta_{1B}^{FS}(L) + \Delta_{MB}^{FS}(L) ,
\end{equation}
where $\Delta_{1B}^{FS}(L)$ includes contributions from the Hartree
energy, the electron-nuclear Coulomb interaction energy, and the
one-body component of the kinetic energy, while $\Delta_{MB}^{FS}(L)$ is
a many-body term that includes contributions from the
exchange-correlation energy and the many-body part of the kinetic
energy. More precisely, $\Delta_{1B}^{FS}(L)$ may be defined as that
part of the total finite-size error that is also present in a DFT
calculation for the same simulation cell and can be corrected using DFT
results.

In their original paper,\cite{kzk} Kwee, Zhang and Krakauer considered
the finite-size errors affecting a QMC simulation carried out in a
supercell of $L \times L\times L$ primitive unit cells with twist
$\mathbf{k}_s = \mathbf{0}$. The corresponding one-particle finite-size
error is
\begin{equation}
    \Delta_{1B}^{FS}(L) = E^{DFT}(\infty) - E^{DFT}(L),
\end{equation}
where $E^{DFT}(\infty)$ is the DFT energy obtained using a fully
converged $k$-point mesh and $E^{DFT}(L)$ is the $\Gamma$-point DFT
energy of the supercell.  This, of course, can be calculated using an
$L \times L \times L$ Monkhorst-Pack grid of $k$ points in the primitive
Brillouin zone. Since $\mathbf{k}_s = \mathbf{0}$, the Monkhorst-Pack
grid includes the origin. The KZK approximation to the many-body
finite-size error is
\begin{equation}
  \Delta_{MB}^{FS}(L) \approx E^{DFT}(L) - E^{KZK}(L),
\end{equation}
where $E^{KZK}(L)$ is the $\Gamma$-point DFT energy of the supercell
computed using the KZK functional instead of the standard LDA. The KZK
approximation to the total finite-size error is
\begin{equation}
  \Delta E^{FS}(L) = \Delta^{FS}_{1B}(L) + \Delta^{FS}_{MB}(L) 
  \approx E^{DFT}(\infty) - E^{KZK}(L).
\end{equation}

If twist averaging is used, the DMC energy becomes a function of the
twist $\mathbf{k}_s$, which lies in the small Brillouin zone
corresponding to the simulation supercell.  The Slater determinants
appearing in the twisted trial wavefunction are built using orbitals
from a Monkhorst-Pack grid of $L \times L \times L$ $k$ points within
the larger primitive Brillouin zone, offset by $\mathbf{k}_s$ from the
origin.  In a DFT context, carrying out a supercell calculation at a
non-zero twist $\mathbf{k}_s$ is equivalent to approximating the
integration over the primitive Brillouin zone by a quadrature over this
offset grid of $k$ points.

To help analyze the FS errors, we write the DMC ground-state energy of
the infinite simulation cell as
\begin{equation}
  E^{DMC}(\infty) \approx E^{DMC}_{TAV}(L) + \Delta E^{FS}_{TAV}(L),
  \label{eq14}
\end{equation}
where
\begin{equation}
  E^{DMC}_{TAV}(L) = \frac {1}{N_{\rm twist}} \sum_{\mathbf{k}_s} 
  E^{DMC}(L,\mathbf{k}_s) 
\end{equation}
is the twist-averaged DMC energy of the $L \times L \times L$ simulation
cell and $\Delta E^{FS}_{TAV}(L)$ is the required FS correction. In the
spirit of KZK, this is approximated using the formula
\begin{equation}
  \Delta E^{FS}_{TAV}(L) \approx E^{DFT}(\infty) - E^{KZK}_{TAV}(L),
\end{equation}
where $E^{DFT}(\infty)$ is the DFT energy computed within the LDA using
a fully converged $k$-point mesh (which in this work means
$28\times28\times28$) and
\begin{equation}
  E^{KZK}_{TAV}(L) = \frac {1}{N_{\rm twist}} \sum_{\mathbf{k}_s}
  E^{KZK}(L,\mathbf{k}_s)
  \label{eq15}
\end{equation}
is the twist-averaged KZK energy for the supercell, computed using the
same set of $N_{\rm twist}$ twists employed in the DMC simulations.  The
FS correction, $\Delta E^{FS}_{TAV}(L)$, accounts both for the many-body
FS errors and for any one-body FS errors not removed by the limited
twist averaging employed in the DMC simulations.

The use of twisted boundary conditions requires the use of complex trial
wavefunctions and increases the computational cost a little because
complex arithmetic is slower than real arithmetic. In the VMC and
wavefunction optimization algorithms, since the expectation values of
Hermitian operators must be real, only the real parts of the
local-energy components need to be calculated and collected. The
run-time and programming-time costs of twist averaging are therefore
small. The use of a complex trial function in DMC requires the
replacement of the fixed-node approximation, in which the DMC
wavefunction is constrained to have the same sign as the trial
wavefunction, by the fixed-phase approximation,\cite{ortiz} in which the
DMC and trial wavefunctions are constrained to have the same phase. In
practice, however, the fixed-node and fixed-phase algorithms are very
similar and little extra coding is required: the real part of the drift
vector is used when proposing trial electron moves; it is neither
necessary nor possible to reject node-crossing electron moves; and, as
in VMC, only the real parts of the local energies are gathered. Another
important issue in twist averaging is the Jastrow factor. This work uses
the same optimized Jastrow for each twist vector, as we found that
re-optimizing the Jastrow factor at every twist provides negligible
improvements in the final results.\cite{backflow} We note, finally, that
the VMC or DMC runs at each twist can be relatively short and need not
be fully converged. The idea is that we collect enough data to achieve
an acceptable error bar when the data are averaged over all twist
vectors. If a normal run without twist averaging takes $N$ moves to
arrive at an acceptable error bar, each twist angle need only be run for
around $N/N_{\rm twist}$ moves. 

\section {Results and discussion}

This section presents DMC results obtained using the three different
FS-correction methods explained in the Sec.~\ref{subsec:fse}.  As simple
example metals, we have studied lithium (Li) and aluminum (Al), with one
and three valence electrons, respectively. The frozen ionic cores are
represented by non-local norm-conserving LDA pseudopotentials.  The KZK
functional is essentially an LDA, so the use of LDA pseudopotentials
allows us to obtain a consistent comparison of all three
finite-size-correction approaches considered. There is evidence that
Hartree-Fock pseudopotentials may produce more accurate results than LDA
pseudopotentials when used in DMC simulations, but since our aim is to
investigate FS errors, and since these are almost independent of
pseudopotential, no advantage would be gained by using another
pseudopotential type.  To check the accuracy and convergence of the DMC
energy as a function of the number of atoms $N$ in the simulation cell,
we have performed calculations for a range of different values of $N$
(and thus also different values of $L$).

We first investigate the effects of applying Chiesa and MPC corrections
to the results of $\Gamma$-point DMC simulations with no twist
averaging.  Table \ref{GAMLi} shows the $\Gamma$-point DMC results for
Li obtained using various different FS correction methods. As
expected,\cite{Neil} results (denoted Ewald + $\Delta$KE + $\Delta$PE)
obtained by adding Chiesa kinetic and potential energy corrections to
the DMC energy calculated using the Ewald interaction are in good
agreement with results (denoted MPC + $\Delta$KE) obtained using the
modified periodic Coulomb interaction with a Chiesa correction for the
kinetic energy only. Note that the higher-order kinetic energy corrections
defined according to equation (55) in Ref.~\onlinecite{Neil}, are included 
in $\Delta$KE. Because of the lack of twist averaging, the
single-body FS errors are large and the calculated ground-state energies
depend strongly on the size of the simulation cell. The choice of the
$\Gamma$-point, ${\bf k}_s = 0$, maintains the symmetry of the system
but usually increases shell-filling effects in metallic systems, making
the independent-particle FS errors even worse. It is clear, however,
that no calculation carried out at a single twist vector will yield
satisfactorily accurate results. The use of twist averaging is essential
in metals.

\begin{table}
  \caption{
    DMC energies of metallic lithium for different numbers $N$ of 
    atoms in the simulation cell. Every energy appearing in the 
    table is the outcome of a single $\Gamma$-point DMC simulation 
    for the appropriate simulation cell. 
    Results obtained by applying Chiesa kinetic ($\Delta$KE) and 
    potential ($\Delta$PE) finite-size corrections to DMC energies 
    calculated using the Ewald interaction agree well with results 
    obtained by applying only the Chiesa kinetic correction to DMC 
    energies calculated using the modified periodic Coulomb 
    interaction.  
    Despite the application of many-body finite-size corrections,
    the calculated energy depends strongly on the size of the 
    simulation cell. This indicates that the single-particle 
    finite-size errors are large. Energies are in eV per atom.
  }
\label{GAMLi}  
\begin{center}
\begin{tabular}{c c c c c c c}
\hline
 N & Ewald  & Ewald + $\Delta$KE & Ewald + $\Delta$PE & Ewald + $\Delta$KE + $\Delta$PE & MPC & MPC + $\Delta$KE \\
\hline\hline
 32 & -6.94307(3) & -6.92246(3) & -6.87450(3) & -6.85389(3) & -6.86219(4) & -6.84157(4) \\
 48 & -6.97920(7) & -6.97852(7) & -6.91729(7) & -6.91661(7) & -6.91661(6) & -6.91593(6) \\
 72 & -6.98253(3) & -6.97981(3) & -6.94471(3) & -6.94205(3) & -6.94062(3) & -6.93790(3) \\
 96 & -6.97063(5) & -6.96845(5) & -6.94015(5) & -6.93804(5) & -6.94382(5) & -6.94171(5) \\
 144& -6.89491(2) & -6.89089(2) & -6.87634(2) & -6.87232(2) & -6.87640(2) & -6.87246(2) \\
\hline\hline
\end{tabular}
\end{center}
\end{table}

Tables \ref{TAVLi} and \ref{TAVAl} present twist-averaged DMC results
for Li and Al, respectively, again corrected using the Chiesa and MPC
approaches.  The integration over the simulation-cell Brillouin zone
that produces the twist-averaged energy would completely remove the
single-particle FS errors if carried out exactly, but in practice the
integration has to be approximated as a summation over a finite set of
twists. The summation tends to the integral as either the number $N$ of
atoms in the simulation cell or the number $N_{\rm twist}$ of twists
used tends to infinity, but is far from perfect in practice. Checking
the convergence with respect to both $N$ and $N_{\rm twist}$ is
important, since fully twist-averaged calculations for finite simulation
cells still contain many-body FS errors and do not necessarily provide
accurate results. The main reason is that the fully twist-averaged
exchange-correlation energy still depends on $N$, even though the
one-electron part of the fully twist-averaged kinetic energy does not.
Comparing the finite-size-corrected and twist-averaged DMC energies from
Tables \ref{TAVLi} with the $\Gamma$-point energies from Table
\ref{GAMLi} shows that twist averaging much reduces the finite-size
errors and allows accurate results to be obtained using much smaller
simulation cells. In this particular case (but not in general) it also
produces a lower ground state energy.

Our twist-averaged DMC results were obtained using $24$ randomly sampled
twists (values of $\mathbf{k}_s$) in the simulation-cell Brillouin
zone. Two other practical sampling schemes exist. One is to use a
uniform Monkhorst-Pack grid \cite{Monkhorst} of $\mathbf{k}_s$ points
centered on the $\Gamma$-point of the simulation-cell Brillouin zone,
and the other is to use a uniform grid centered on the Baldereschi point
\cite{Balder} of the simulation-cell Brillouin zone. As either the
number of twists or the size of the simulation cell tends to infinity,
all three twist-averaging methods yield the same results.

\begin{table}
  \caption{ 
    DMC energies of lithium for different numbers $N$ of atoms in 
    the simulation cell. Every energy appearing in the table is an 
    average of the outcomes of $24$ separate DMC simulations with
    different randomly chosen twists. Results obtained by applying 
    Chiesa kinetic ($\Delta$KE) and potential ($\Delta$PE) finite-size
    corrections to DMC energies calculated using the Ewald 
    interaction agree well with results obtained by applying only 
    the Chiesa kinetic correction to DMC energies calculated 
    using the modified periodic Coulomb interaction. The use of twist
    averaging has much reduced the independent-particle finite-size
    errors observed in Table \ref{GAMLi}. Energies are in eV per 
    atom.
  }
\label{TAVLi}  
\begin{center}
\begin{tabular}{ c c c c c c c }
\hline
 N & Ewald  & Ewald + $\Delta$KE & Ewald + $\Delta$PE & Ewald + $\Delta$KE + $\Delta$PE & MPC & MPC + $\Delta$KE \\
\hline\hline
 32 & -7.0008(4) & -6.9802(4) & -6.9334(4) & -6.9128(4) & -6.9301(6) & -6.9095(6) \\
 48 & -6.9708(4) & -6.9699(4) & -6.9100(4) & -6.9092(4) & -6.9089(5) & -6.9080(5) \\
 72 & -6.9590(3) & -6.9563(3) & -6.9207(3) & -6.9180(3) & -6.9203(2) & -6.9177(2) \\
 96 & -6.9489(2) & -6.9467(2) & -6.9184(2) & -6.9162(2) & -6.9207(2) & -6.9176(2) \\
 144& -6.9291(2) & -6.9251(2) & -6.9113(2) & -6.9073(2) & -6.9118(2) & -6.9078(2) \\ 
\hline\hline
\end{tabular}
\end{center}
\end{table}

In applications of DMC to real systems using computers routinely
available to researchers, it is rarely possible to treat very large
simulation cells or numbers of twists. Restricting the number of twists
is particularly problematic in metallic systems, where the Fermi surface
discontinuity makes the integrand (for example, the total kinetic
energy) a discontinuous function of $\mathbf{k}_s$ at zero temperature.
The convergence with system size and number of twists is therefore much
slower for metals than for insulators. Hartree-Fock calculations for the
uniform electron gas\cite{Neil} show that energies obtained using
Baldereschi twist averaging converge faster than energies obtained using
random twist averaging at very large numbers of twists (although it
could be argued that the choice of the Baldereschi point introduces a
systematic bias into the unconverged results), but that both methods
converge slowly. Here we show that the use of the
incomplete-twist-averaging correction defined in Eq.~(\ref{deltaBZ})
allows well-converged results to be obtained with very small numbers of
twists. The choice between Baldereschi or random twist sampling is then
unimportant.

\begin{table}
  \caption{ 
    DMC energies of aluminum for different numbers $N$ of atoms in 
    the simulation cell. Every energy appearing in the table is an 
    average of the outcomes of $24$ separate DMC simulations with
    different randomly chosen twists. Results obtained by applying 
    Chiesa kinetic ($\Delta$KE) and potential ($\Delta$PE) finite-size
    corrections to DMC energies calculated using the Ewald 
    interaction agree well with results obtained by applying only 
    the Chiesa kinetic correction to DMC energies calculated 
    using the modified periodic Coulomb interaction. Energies 
    are in eV per atom.
  }
\label{TAVAl}  
\begin{center}
\begin{tabular}{ c c c c c c c }
\hline
 N & Ewald  & Ewald + $\Delta$KE & Ewald + $\Delta$PE & Ewald + $\Delta$KE + $\Delta$PE & MPC & MPC + $\Delta$KE \\
\hline\hline
 24 & -56.175(2)  & -56.170(2)  & -55.932(2)  & -55.927(2)  & -56.095(2)  & -56.091(2) \\
 32 & -56.203(1)  & -56.152(1)  & -56.074(1)  & -56.022(1)  & -56.0585(2)  & -56.003(2) \\
 48 & -56.155(1)  & -56.138(1)  & -56.048(1)  & -56.031(1)  & -56.058(2)  & -56.041(2) \\ 
 72 & -56.0922(7) & -56.0808(7) & -56.0212(7) & -56.0098(7) & -56.0226(8) & -56.0112(8) \\
\hline\hline
\end{tabular}
\end{center}
\end{table}  

Tables \ref{CorrectLi} and \ref{CorrectAl} compare twist-averaged DMC
results obtained using the $\Delta$BZ incomplete-twist-averaging FS
correction and several different many-body FS-correction methods.  The
convergence of the FS-corrected DMC energies with system size $N$ is
excellent and there is no difficulty in reaching an accuracy of a few
meV per atom. As before, our DMC energies are averages over $24$
randomly chosen twists in the simulation-cell Brillouin zone,
corresponding to $24$ randomly translated $L \times L \times L$
Monkhorst-Pack $k$-point meshes in the primitive Brillouin zone.  We
also carried out twist-averaged DMC calculations using $36$ twists; the
change in the total energy was less than 1.5 meV/atom for Li and less
than 2.7 meV/atom for Al. This shows that twist-averaged DMC energies
including $\Delta$BZ corrections converge very rapidly as the number of
twists is increased.

\begin{table}
  \caption{ 
    DMC energies of lithium for different numbers $N$ of atoms in 
    the simulation cell. Every DMC energy is an average of the 
    outcomes of $24$ separate DMC simulations with different 
    randomly chosen twists. The incomplete-twist-averaging
    finite-size correction $\Delta$BZ is included in all energies.
    Results obtained by applying the Chiesa kinetic ($\Delta$KE) 
    and potential ($\Delta$PE) finite-size corrections to DMC 
    energies calculated using the Ewald interaction agree well 
    with results obtained by applying the Chiesa kinetic correction 
    to DMC energies calculated using the modified peiodic Coulomb 
    interaction. Results obtained using the twist-averaged KZK 
    method, which also include an equivalent of the $\Delta$BZ 
    correction, are also in good agreement. Energies are in eV 
    per atom.
  }
\label{CorrectLi}  
\begin{center}
\begin{tabular}{ c c c c }
\hline
 N & Ewald + $\Delta$KE + $\Delta$PE + $\Delta$BZ & MPC + $\Delta$KE + $\Delta$BZ  & TAV-KZK \\
\hline\hline
 32 & -6.9299(4) & -6.9267(4) & -6.9126(4) \\
 48 & -6.9183(4) & -6.9172(4) & -6.9095(4) \\
 72 & -6.9186(3) & -6.9182(3) & -6.9126(3) \\
 96 & -6.9149(2) & -6.9163(2) & -6.9123(2) \\
144 & -6.9142(2) & -6.9147(2) & -6.9125(2) \\ 
\hline\hline
\end{tabular}
\end{center}
\end{table}

\begin{table}
  \caption{
    DMC energies of aluminum for different numbers $N$ of atoms in 
    the simulation cell. Every DMC energy is an average of the 
    outcomes of $24$ separate DMC simulations with different 
    randomly chosen twists. The incomplete-twist-averaging
    finite-size correction $\Delta$BZ is included in all results.
    Results obtained by applying the Chiesa kinetic ($\Delta$KE) 
    and potential ($\Delta$PE) finite-size corrections to DMC 
    energies calculated using the Ewald interaction agree well 
    with results obtained by applying the Chiesa kinetic correction 
    to DMC energies calculated using the modified peiodic Coulomb 
    interaction. Results obtained using the twist-averaged KZK 
    method, which also include an equivalent of the $\Delta$BZ 
    correction, are also in good agreement. Energies are in eV 
    per atom.
 }
\label{CorrectAl}  
\begin{center}
\begin{tabular}{ c c c c }
\hline
 N & Ewald + $\Delta$KE + $\Delta$PE + $\Delta$BZ & MPC + $\Delta$KE + $\Delta$BZ  & TAV-KZK \\
\hline\hline
 24 & -56.019(2)  & -56.183(2)  & -56.025(2) \\
 32 & -56.088(1)  & -56.069(2)  & -56.088(1) \\
 48 & -56.078(1)  & -56.088(2)  & -56.081(1) \\ 
 72 & -56.0607(7) & -56.0621(8) & -56.0626(7) \\
\hline\hline
\end{tabular}
\end{center}
\end{table}

\begin{figure}
\includegraphics{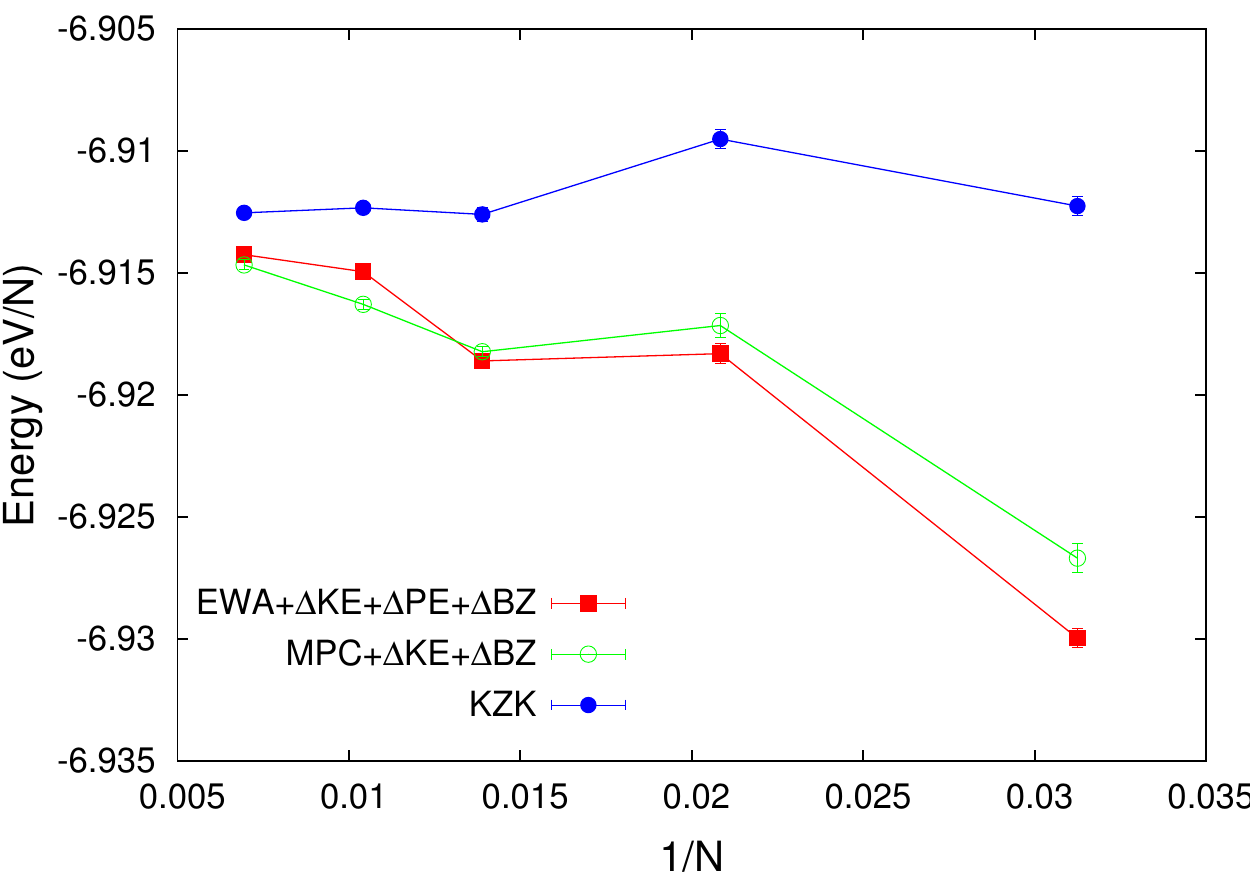}
\caption{\label{LiKZK} Total DMC energies of Li as function of the
  number of particles in the simulation cell. All energies are averaged
  over 24 randomly chosen twists and include $\Delta$BZ corrections. Red
  squares indicate DMC results calculated using the Ewald interaction
  with Chiesa corrections for the kinetic and potential energies (Ewald
  + $\Delta$KE + $\Delta$PE + $\Delta$BZ); green circles indicate DMC
  results obtained using the modified periodic Coulomb interaction with
  Chiesa corrections for the kinetic energy only (MPC + $\Delta$KE +
  $\Delta$BZ); and blue circles indicate KZK-corrected DMC results,
  again incorporating $\Delta$BZ corrections.}
\end{figure}

Figure \ref{LiKZK} shows how the FS-corrected DMC energies of metallic
Li depend on system size, allowing an easy comparison of the three
different many-body FS-correction methods considered in this work. All
DMC energies are averaged over 24 randomly chosen twists and include
$\Delta$BZ corrections. Red squares indicate DMC results calculated
using the Ewald interaction with Chiesa corrections for the kinetic and
potential energies (Ewald + $\Delta$KE + $\Delta$PE + $\Delta$BZ); green
circles indicate DMC results obtained using the modified periodic
Coulomb interaction with Chiesa corrections for the kinetic energy only
(MPC + $\Delta$KE + $\Delta$BZ); and blue circles indicate KZK-corrected
DMC results, again incorporating $\Delta$BZ corrections.  Even for the
smallest simulation cell considered, with just 32 atoms, the errors in
the Ewald + $\Delta$KE + $\Delta$PE + $\Delta$BZ and MPC + $\Delta$KE +
$\Delta$BZ total energies are only 15.7 and 12 meV/atom,
respectively. The errors in Ewald DMC energies corrected using the KZK
scheme are even smaller, at approximately 3 meV/atom.

We emphasize the importance of the success of the twist-averaged KZK
method from a practical point of view. It is known, for example, that
the cheap and widely used DFT approach often fails to provide accurate
enough results\cite{dft-fail} to understand the behavior of materials at
high pressure. Therefore, to study the very interesting phase diagram of
Li,\cite{HPLi} it will be necessary to perform full many-body
computations, most likely using DMC. The drawback is that DMC
calculations are typically at least a thousand times more expensive than
DFT calculations. The twist-averaged KZK approach allows one to
investigate a large number of possible crystal structures and construct
the Li phase diagram whilst keeping the cost of the DMC simulations
within reasonable bounds.

\begin{figure}
\includegraphics{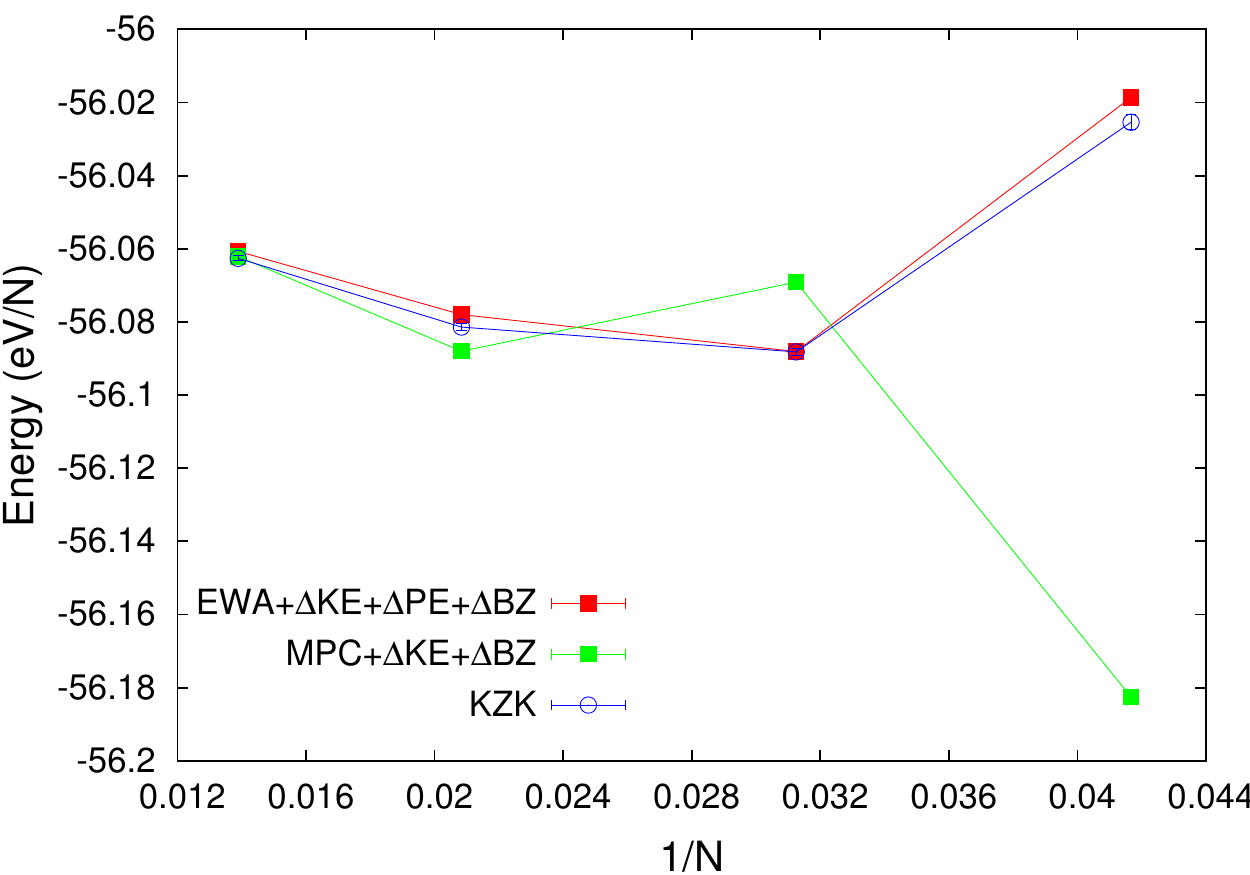}
\caption{\label{AlKZK} Total DMC energies of Al as function of the
  number of particles in simulation cell. All energies are averaged over
  24 randomly chosen twists and include $\Delta$BZ corrections. Red
  squares indicate DMC results calculated using the Ewald interaction
  with Chiesa corrections for the kinetic and potential energies (Ewald
  + $\Delta$KE + $\Delta$PE + $\Delta$BZ); green circles indicate DMC
  results obtained using the modified periodic Coulomb interaction with
  Chiesa corrections for the kinetic energy only (MPC + $\Delta$KE +
  $\Delta$BZ); and blue circles indicate KZK-corrected DMC results,
  again incorporating $\Delta$BZ corrections.}
\end{figure}

Figure \ref{AlKZK} shows how the FS-corrected DMC energy of metallic Al
depends on the number of atoms in the simulation cell. All DMC energies
are averaged over $24$ randomly chosen twists and include $\Delta$BZ
corrections.  Red squares indicate FS-corrected results obtained using
the Ewald interaction with Chiesa corrections for the kinetic and
potential energies (Ewald + $\Delta$KE + $\Delta$PE + $\Delta$BZ); green
circles indicate DMC results obtained using the modified periodic
Coulomb interaction (MPC) with Chiesa corrections for the kinetic energy
only (MPC + $\Delta$KE + $\Delta$BZ); and blue circles indicate
KZK-corrected DMC results, again incorporating $\Delta$BZ corrections.
For all simulation-cell sizes, the Ewald + $\Delta$KE + $\Delta$PE +
$\Delta$BZ and KZK results are in almost perfect agreement.  The
difference between the MPC + $\Delta$KE + $\Delta$BZ energies and those
obtained using the other two FS-correction methods is about 7 meV/atom
for a cell containing just 48 atoms and decreases rapidly with
increasing simulation-cell size.

The results of this section have shown that the addition of an
incomplete twist-averaging correction, $\Delta$BZ, allows accurate
results to be obtained with a remarkably small number of twists, even
for metals. In most of the cases studied (except for the 96-atom Li
simulation cell), the $\Delta$BZ correction lowers the total DMC
energy. The values of $\Delta BZ$ for Li simulation cells containing 32,
48, 72, 96, and 144 atoms are -0.0171, -0.0091, -0.0006, +0.0013, and
-0.0069 eV, respectively. In the case of Al simulation cells containing
24, 32, 48, and 72 atoms, the values of $\Delta$BZ are -0.092, -0.066,
-0.047, and -0.050 eV, respectively.
 A recent paper by Shulenberger and
Mattsson \cite{shulenberger2013} provided accurate benchmark DMC results
for a wide range of different bulk materials.  They required 216 and 64
twists to obtain converged results for Li and Al supercells containing
28 and 108 atoms, respectively. Because we use incomplete
twist-averaging corrections, $\Delta$BZ, we are able to obtain similarly
accurate results with considerably fewer twists.

\section {Conclusion}

We have systematically analyzed and compared the various schemes that
have been proposed for correcting FS errors in QMC simulations of real
metallic systems. We have explained how to combine the use of
twist-averaged boundary conditions with the KZK functional and shown the
value of incomplete-twist-averaging corrections based on DFT.  The
reassuring news is that all of the commonly used approaches work well.

We believe that the use of DFT-based incomplete-twist-averaging
corrections will have an important role to play in DMC simulations of
real metallic systems. The reliance on DFT could be considered a
drawback, but it is important to bear in mind that \emph{any} valid
FS-correction method must yield the same total energy in the limit of a
large enough simulation cell. The important question is not whether the
unattainable limiting value is correct but how rapidly it is
approached. The use of incomplete-twist-averaging corrections
significantly improves this convergence. Furthermore, energies
calculated using the twist-averaged KZK scheme (which implicitly
incorporates a $\Delta$BZ incomplete-twist-averaging correction) often
settle down to a system-size-independent constant more quickly than
energies calculated using other methods incorporating $\Delta$BZ
corrections.

We believe that this paper will provide a useful guide and benchmark for
researchers using QMC and other many-body electronic structure methods
such as CCSD(T) and the $GW$ approximation to study metallic systems.

\begin{acknowledgments}
This work was supported by the UK Engineering and
Physical Sciences Research Council under grant EP/K038141/1,
and made use of computing facilities provided by ARCHER,
the UK National Supercomputing Service, and by the High 
Performance Computing Service of Imperial College London. We
acknowledge support from the Thomas Young Centre under
grant TYC-101.
\end{acknowledgments}


\begin{thebibliography}{99}
\bibitem{backflow} In the case of high-pressure solid hydrogen, we have found that re-optimizing the backflow transformation at each twist improves the final results.
\end{thebibliography}
\end{document}